\documentclass[structabstract]{aa}  
\usepackage{graphicx}

\usepackage{natbib}
\usepackage{txfonts}
\begin{document}
\authorrunning{Gorlova et al}
\titlerunning{A new evolved binary with a disk}

   \title{Time resolved spectroscopy of BD+46$\degr$442: gas streams and jet creation in a newly discovered evolved binary with a disk\thanks{Based on observations made with the Mercator Telescope,
operated on the island of La Palma by the Flemish Community, at the Spanish Observatorio
del Roque de los Muchachos of the Instituto de Astrof\'{i}sica de Canarias.}}


   \author{N. Gorlova\inst{1}
          \and H. Van Winckel\inst{1}
          \and C. Gielen\inst{1}
          \and G. Raskin\inst{1}
          \and S. Prins\inst{1}
          \and W. Pessemier\inst{1}
          \and C. Waelkens\inst{1}
          \and Y. Fr\'{e}mat\inst{2}
          \and H. Hensberge\inst{2}
          \and L. Dumortier\inst{2}
          \and A. Jorissen\inst{3}
          \and S. Van Eck\inst{3}
          }

   \institute{Instituut voor Sterrenkunde, Katholieke Universiteit Leuven,
              Celestijnenlaan 200D, 3001 Leuven, Belgium\\
              \email{nadya@ster.kuleuven.be}
          \and
             Royal Observatory of Belgium, 3 Avenue Circulaire, 1180 Brussels, Belgium
          \and
             Institut d'Astronomie et d'Astrophysique, Universit\'{e} Libre de Bruxelles,
             CP 226, Boulevard du Triomphe, 1050, Bruxelles, Belgium
             }


 
  \abstract
   {Previous studies have shown that many post-AGB stars with dusty disks are
    associated with single-lined binary stars. The inferred orbital separations are too small
    to accommodate a fully-grown AGB star,
    therefore, these systems represent a new
    evolutionary channel that bypasses a full AGB evolution.}
   {To verify the binarity hypothesis on a larger sample, to reveal the nature of the companions,
    and to probe the disk structure and eventually the disk formation mechanisms, we started
    a high-resolution spectral monitoring of $\sim$40 field giants
    whose binarity was suspected based on either a light curve, an infrared excess,
    or a peculiar chemical composition.}
   {Starting from the spring of 2009, we monitor the program stars
    with a new fiber echelle spectrometer HERMES.
    We measure their radial velocities (RVs) with a precision of $\sim$0.2 km~s$^{\mathrm{-1}}$, perform
    detailed photospheric abundance analyses, and analyze the
    time-resolved high-resolution spectra to search for line-profile variability.}
   {Here we report on the discovery of the periodic RV variations in BD+46$\degr$442,
    a high-latitude F giant with a disk. We interpret the variations due to the motion
    around a faint companion, and deduce the following orbital parameters:
    $P_{\mathrm{orb}} = 140.77 \pm 0.02\, ^{\rm d}, e = 0.083 \pm 0.002, a\sin i=0.31$ AU.
    We find it to be a moderately metal-poor star ($\mathrm{[M/H]}=-0.7$)
    without a strong depletion pattern in the photospheric abundances.
    Interestingly, many lines indeed show periodic changes with the orbital phase:
    H$\alpha$ switches between a double-peak emission and a P~Cyg-like profiles,
    while strong metal lines appear split during the maximum redshift.
    Similar effects are likely visible in the spectra of other post-AGB binaries,
    but their regularity is not always realized due to sporadic observations.
    We propose that these features result from an ongoing mass transfer
    from the evolved giant to the companion. In particular, the blue-shifted
    absorption in H$\alpha$, which occurs only at superior conjunction,
    may result from a jet originating in the accretion disk around the companion and seen in
    absorption towards the luminous primary.} 
   {}


   \maketitle
%

\section{Introduction}

Recently \citet{VanArle2011} investigated the spectral energy distribution (SED)
of optically bright candidate post-asymptotic giant branch (post-AGB) stars in the Large Magellanic Cloud.
Surprisingly, they found that in about half of candidates, the SED
indicates that the circumstellar material, which is the relic of the
AGB mass loss, is not in the form an exanding envelope, but rather in
the form of a stable, likely Keplerian, disk.

In our own Galaxy, post-AGB stars with disks are common as well \citep[e.g.][]{Deruyter2006},
but statistics of Galactic post-AGB stars are less complete, and the lack of
well constrained distances (hence, luminosity) makes the interpretation
of these numbers more difficult. Galactic sources, however, have an obvious advantage of being closer
and easier to study.

The leading hypothesis explaining the envelope bifurcation into the outflows and the disk types,
is that disks form only around binaries. The observational foundation
comes from the fact that indeed, many disc sources in our Galaxy turned out
to be single-lined binaries \citep[e.g.][and references therein]{VanWinckel2009}.
Presumably disk formation happens through an interaction with the companion when the more massive component
becomes a red giant (RG) or an asymptotic giant branch (AGB) star
\citep{VanWinckel2003}. The process itself is poorly understood.
\citet{Livio1988} and \citet{Sandquist1998}, among others,
showed that a circumbinary disk could be a remnant of a common envelope
created when the Roche lobe of an AGB star engulfs a companion. Other
scenarios were proposed by e.g. \citet{Mastrodemos1999}, where a
disk-like structure is created in a wind-accretion scenario
or perhaps as a result of the AGB wind and the accretion jets
interaction\citep{Akashi2008}. 

The impact of the disk onto the system cannot be underestimated.
Many of the Galactic and even LMC post-AGB disk objects show a
peculiar chemical photospheric
composition that resembles the gas phase of the interstellar medium
(ISM): elements of higher condensation temperature, such as iron, are
less abundant than volatiles
\citep[e.g.][]{Giridhar2005,Maas2005,Gielen2009,Rao2011}.  This is believed to be
a result of pollution from a disk, which may have already dissipated in some systems and
therefore has no footprint in the SED.
In the disk, refractory elements nucleate into dust grains, which are
segregated from the gas, and the depleted gas falls back onto the star
\citep{Waters1992}. On one hand, this effect helps to identify
current or former disk systems, but on the other hand masks the original metallicity of the
stars and the eventual products of the AGB dredge-ups, complicating the
identification of the evolutionary stage.
 
At a later evolutionary stage, as was proposed by \citet{Waelkens1996} and \citet{Jorissen1999},
a disk system may turn into a Barium star,
when the current post-AGB primary evolves into a white dwarf (WD) and supposedly
a main sequence (MS) companion into a red giant.
Barium stars are binaries where a red giant bears signs of contamination
by the nuclear-synthetic products (such as Barium) by the former AGB companion (now a WD).
By adopting the formalism of \citet[][and references therein]{Lubow2010},
\citet{Dermine2010} explored the binary-disk interaction regime
relevant in Barium stars and showed that circumbinary disks
may be important at pumping the eccentricities of Barium stars to the
observed values.

The Galactic objects with known orbits cannot have evolved on
single-star evolutionary tracks: typical AGB star has a radius $\sim$1 AU,
while many post-AGB binaries with a disk have smaller separations,
which brings a possibility that these systems started interacting
already at the red giant stage. Some of the objects may have never gone through the AGB stage,
unless the interaction resulted in the orbit shrinkage. 
With the
observed periods, the full AGB
evolution must be shortcut and this is corroborated by the finding that
the majority of disk sources have oxygen-rich photospheres and disks dominated
by crystalline silicates \citep{Gielen2008, Gielen2011}.
Another ambiguity pertaining the evolution of these systems
is whether the stars will still show up as planetary nebulae (PNe), with the circumstellar
material locked in a disk. 
While binarity helps to explain bipolar PNe,
the periods of binary stars in PNe are at least two orders of magnitude shorter
than in the disk post-AGB systems \citep{DeMarco2009}. 

The theories of disk formation cannot
predict timescales for disk re-accretion and dissipation either.
The insight can be gained by gathering information
on the inner-most regions of the disk from optical and near-IR spectroscopy.
Previous spectroscopic studies of post-AGB stars explored mainly stellar radial velocities (RVs)
and the basic properties of both the photosphere and the disk.
Only briefly did they touch on the signs of activity
at the star-to-disk interface, such as emission in H$\alpha$ or circumstellar
absorptions in the \ion{Ca}{II} \rm{H\&K} and \ion{Na}{I} \rm{D} lines,
and the interpretation was usually that these features trace a subsiding AGB wind.
The notion was that after a circumbinary disk is formed, the system
enters a passive evolution as a detached binary.
No success has been achieved at uncovering gas re-accretion
\citep[with only one tentative detection of the winnowing gas by][]{Hinkle2007}.
In none of the systems, the companion is detected so far.
To explore the details of the disk-star interaction, systematic
time-resolved investigation 
of the disk and chemically-peculiar post-AGB systems is therefore
highly desirable. 

A literature compilation of Galactic post-AGBs
was assembled in the Toru\'{n} catalog \citep{Szczerba2007}\footnote{http://www.ncac.torun.pl/postagb2}
and consists of $\sim$400 objects.
In this catalog, 51 source have been identified as disk sources in \citet {Deruyter2006}, and
this number has increased ever since.
About two dozen post-AGBs of De Ruyter et al's list have been identified to be single-lined spectroscopic
binaries with periods from 115 to 2600 days and orbital separations
from 0.1 to 3.6 AU. The latest addition can be found in \citet{VanWinckel2009}
who photometrically and spectroscopically monitored a selection of
low-amplitude pulsators from the southern hemisphere. All
low-amplitude pulsators are binaries, but the suspected binary nature
of the remaining 60\% of De Ruyter et al's list remains to be
established. This is an observational challenge, because many objects
show large amplitude pulsations making the interpretation of radial
velocity variations not straightforward \citep[e.g.][]{Maas2002}.

In the spring of 2009, we started a high-resolution spectroscopic
monitoring programme on 34 Galactic post-AGB candidates with disks and a few objects
without the IR excess, that are observable from the northern hemisphere,
using a newly built optical echelle spectrograph HERMES \citep{Raskin2011}.
These objects were required to satisfy at least two
of the following criteria: an IR excess characteristic of a disk,
an RV Tau-like variability, variability suggesting a possible
obscuration by a disk, low surface gravity combined with the location away from the Galactic plane,
deficiency in the refractory elements.
In this paper we report on one of the first new short-period binaries
that we uncovered after three years of observations.
BD+46$\degr$442 (SAO 37487, IRAS 01427+4633) is a poorly-studied $V=9.5$
high-latitude Galactic F giant\footnote{When referring to BD+46$\degr$442-like systems,
the term ``giant'' is used to indicate a low surface-gravity,
extended, post-MS object. The exact luminosity class and the evolutionary stage
can only be established when the parallax information becomes available.}
($b=15\degr$), not known to be variable in RV or photometry.
It was selected in our survey due to the strong infrared excess characteristic for disk objects,
as originally measured by the IRAS satellite.
The paper is organized as follows: after description of the data (Sect. \ref{sec_data}),
we present results of the abundance analyses of BD+46$\degr$442 (Sect. \ref{sec_abunds}),
then model its SED (Sect. \ref{sec_sed}), the radial velocity (RV) curve (Sect.  \ref{sec_rv}),
and examine the behaviour of the hydrogen and some metal lines (Sect.  \ref{profs}).
In the Discussion section, we compare the discovered periodic spectral variations
with those in other post-AGB systems and in interacting binaries,
and discuss theories capable to explain this behaviour.


\section{Observations and data reduction}\label{sec_data}

Each target in our on-going post-AGB survey is observed about twice per month during the visibility season.
This program is part of a much broader effort to study binary
interaction processes in evolved stars \citep{VanWinckel2011}.
The main goal is to measure RVs to detect and characterize the orbital motion.
For brighter targets, additional studies are performed, such as
determination of abundances, rotation velocities, and the circumstellar features.
The spectra are obtained with the fiber-fed echelle spectrograph HERMES,
attached to the Flemish 1.2 m telescope Mercator on La Palma, Canary Islands.
The spectrograph is optimized for high resolution, stability, and
broad wavelength coverage. This is achieved primarily by implementing
an image slicer, an anti-fringe CCD coating, and a thermal enclosure
\citep{Raskin2011}.
  
Sixty spectra of BD+46$\degr$442 were obtained during three seasons
of observations in 2009 -- 2012.
We used a fiber configuration that provides a resolution $R= 80,000 - 90,000$ over
the wavelength range $\Delta \lambda = 3800 - 9000$\AA.
The exposure times varied between 500 -- 1800 s while the signal-to-noise ratio (S/N) between 20 -- 90
(as measured near 6500 \AA).
Nightly calibrations consisted of two sets of biases, flats,
and the \rm{Th-Ne-Ar} arc spectra obtained in the evening and in the morning.
To reduce a particular object frame, the nearest set was used.
At least one IAU radial velocity standard was observed nightly
and occasionally a series of arcs,
to monitor the fluctuations in the zero-point of the wavelength calibration.
These tests showed that the typical uncertainty on the absolute value of the RV
is $\sim$0.2 km~s$^{\mathrm{-1}}$ and is mainly caused by the pressure fluctuations
in the instrument room \citep{Raskin2011}.
The data reduction was performed with a dedicated Python-based pipeline,
that outputs extracted, cosmic-ray cleaned, wavelength calibrated, 
and order-merged spectra.
 
\section{Atmospheric analysis}\label{sec_abunds}

\subsection{Atomic data and model atmospheres}

Our atmospheric analysis of BD+46$\degr$442 is based on the comparison with synthetic
profiles of hydrogen lines and on the equivalent-width (EW) measurement of metal lines.
The latter were obtained using the DECH20 code by \citet{Galazutdinov1992}\footnote{http://www.gazinur.com/DECH-software.html},
that provides a convenient interface for the continuum normalization, and the EW measurement
by means of a Gaussian fit or a direct integration.
Line identification was carried out using the following on-line resources: Spectroweb stellar atlas
by \citet{Lobel2008, Lobel2011}\footnote{http://spectra.freeshell.org/spectroweb.html} for
$\lambda < 6800$ \AA\,, and the Solar atlas
by \citet{Delbouille1988}\footnote{http://ljr.bagn.obs-mip.fr/observing/spectrum.html}
in combination with the
Atomic line database by P. Van Hoof\footnote{http://www.pa.uky.edu/$\sim$peter/atomic/} for
the longer wavelengths.

The line list used for the abundance determination
was provided by V. Kovtyukh (2010, private communication)
and is mainly based on the version published in
\citet{Kovtyukh1999}. We also follow the procedure
for the determination of the surface gravity ($\log g$)
and the micro-turbulent velocity ($V_\mathrm{tur}$) from the latter work.
Our choice is dictated by the fact that
yellow supergiants studied by \citet{Kovtyukh1999}
spectroscopically are very similar to post-AGB stars.
To convert EWs to abundances,
we use the local thermodynamical equilibrium (LTE) radiative transfer code
MOOG by C. Sneden\footnote{http://www.as.utexas.edu/$\sim$chris/moog.html}
with the ATLAS9 model atmospheres that include the updated opacity distribution functions
of \citet{Castelli2003}\footnote{http://wwwuser.oat.ts.astro.it/castelli/grids.html}.

\citet{Kovtyukh1999} obtained the bulk of their oscillator strengths ($\log gf$-s)
by adjusting the laboratory values in such a way, that they could reproduce the solar abundances
of \citet{Grevesse1996} from the EWs measured in the solar spectrum of \citet{Kurucz1984}.
They used a canonical solar model of 5777/4.438/1.0 ($T_{\mathrm{eff}}/\log g/V_{\mathrm{tur}}$)
from the set of the original ATLAS9 models of \citet{Kurucz1992}
and his WIDTH9 LTE code \footnote{http://wwwuser.oat.ts.astro.it/castelli/sources/width9.html}.
We found that this combination of the older atmospheric models and the code tends
for the lines of a given element to produce systematically slightly different abundances than our
combination of the atmospheric models and code, both for the Sun and BD+46$\degr$442.
Therefore, when giving abundances of BD+46$\degr$442 relative to the Sun (\rm{[X/H]}),
we introduced small corrections (up to $\pm$0.08 dex) in the solar abundances of \citet{Kovtyukh1999}
to compensate for this difference.

\subsection{Photospheric parameters}\label{sec_phot}
 
\begin{figure}
   \centering
     \includegraphics[width=9.0cm]{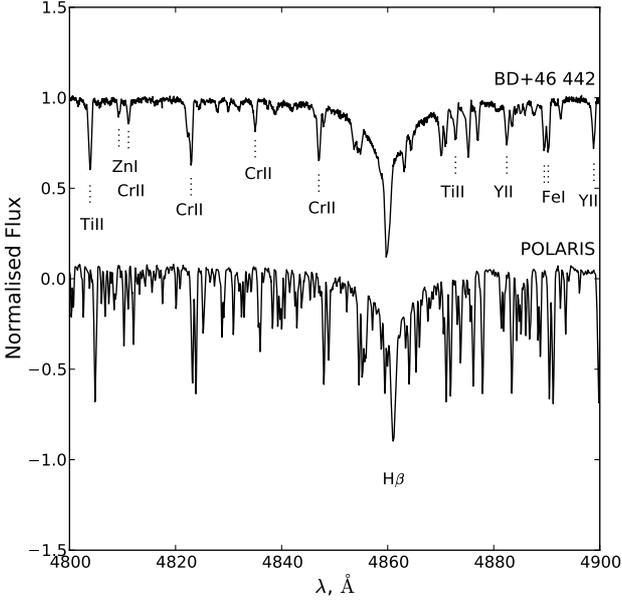}
      \caption{Sample of the spectrum of BD+46$\degr$442 obtained on 2009/08/17 compared to the HERMES spectrum of Polaris.
While H$\beta$ is similar in both stars, indicating similar temperature and gravity,
metal lines are noticeably weaker (and broader) in BD+46$\degr$442, indicating a sub-solar metallicity.
              }
         \label{fig_Pol}
   \end{figure}

Since there is no information on the reddening or
the photometric variability in BD+46$\degr$442,
we decided to derive all photospheric parameters spectroscopically.
For this we analyzed a S/N$\sim$130 spectrum obtained
by averaging the best quality spectra numbered 3 and 4 from the same night (Fig. \ref{fig_Pol}).
The visual examination of all the spectra did not reveal
any obvious variations in the effective temperature ($T_{\mathrm{eff}}$).
Variations in the cores of strong lines have been detected,
which we attribute to the circumstellar matter (Sect. \ref{profs}),
but those lines were not used in the EW analysis
(we only used lines with EWs $\le$ 170 \AA\,).
This single spectrum, therefore, should give a good representation
of the basic photospheric parameters of our star.

First, we confirmed a spectral type F from Simbad by comparing our spectrum to the spectral libraries
of R. O. Gray\footnote{http://ned.ipac.caltech.edu/level5/Gray/Gray$\_$contents.html}
and of the VLT-UVES \citep{Bagnulo2003}\footnote{http://www.sc.eso.org/santiago/uvespop/}.
Then we computed from our EWs the iron abundance, by adopting
a range of model parameters expected for an F I-III star:
$T_{\mathrm{eff}}= 6200-8000$ K, $\log g=0.5-4.0$, and $V_{\mathrm tur}=0-12.0$ km~s$^{\mathrm{-1}}$,
first using the ATLAS9 models with the solar metallicity ($\mathrm{[M/H]}=0$)
and later with $\mathrm{[M/H]}=-0.5$ to match the subsolar metallicity
obtained in the final iteration.
To perform an excitation analysis, 
we plotted the iron abundances against the EW and the excitation potential of the lower level ($\chi$),
separately for the \ion{Fe}{I} and \ion{Fe}{II} lines, to examine possible trends. 
Correct (LTE) atmospheric parameters are obtained when all the lines give the same value of the abundance.
Trends with the EW can be removed by tuning $V_{\mathrm{tur}}$,
while agreement between the average abundances derived from the \ion{Fe}{I} and \ion{Fe}{II} lines
(the ionization balance) indicates a correct $\log g$. 
Iron lines are usually used in this method because they are the most numerous,
but the same values of $V_{\mathrm{tur}}$ and $\log g$ should in principle hold for all the elements.

At sub-solar metallicities and gravities, however, neutral species become underpopulated
compared to the LTE case, due to the UV over-ionization and the decreased rate of collisions,
resulting in the positive corrections that need to be added to the LTE abundances
\citep{Rentzsch-Holm1996, Bergemann2011, Mashonkina2011}.
Observationally this is manifested in the fact
that in supergiants, \ion{Fe}{I} lines require smaller values of $V_{\mathrm{tur}}$ and $\log g$
to bring their abundances in agreement with the better behaved \ion{Fe}{II} lines
(which is the dominant ionization state in F-G stars),
as shown for example in \citet{Kovtyukh1999} for the classical Cepheid $\delta$ Cep
and in \citet{Takeda2007} for some post-AGB stars.
We find the same effect for BD+46$\degr$442.
Following recommendations in the literature, we use the \ion{Fe}{II} lines 
for the determination of $V_{\mathrm{tur}}$, while for the determination of $\log g$
we extrapolate abundances from the \ion{Fe}{I} lines to EW $=$ 0,
as weak lines are less susceptible to non-LTE effects.
Following this procedure, we found a correlation
between the adopted $T_{\mathrm{eff}}$ and the derived $\log g$/[Fe/H].
To break this degeneracy, we turn to the Balmer and Paschen hydrogen lines.

Hydrogen lines have the advantage over the metal lines that to first order they do not
depend on the metallicity.
Furthermore, we can consider the shape of the extended wings
in addition to the line depth when comparing to the synthetic profiles.
We compared Balmer lines in BD+46$\degr$442 to the
synthetic profiles of \citet{Coelho2005}\footnote{http://www.mpa-garching.mpg.de/PUBLICATIONS/DATA/SYNTHSTELLIB/synthetic$\_$stellar$\_$spectra.html} and Paschen lines to those of
\citet{Munari2000}\footnote{http://vizier.u-strasbg.fr/viz-bin/VizieR-4?-source=III/238}.
H$\alpha$ was not considered, as it shows a strong
non-photospheric contribution at all phases.
From the Paschen lines we chose to consider only Pa14 and Pa17 as the least blended ones. 
We noticed that the Paschen break in the models is too deep compared to BD+46$\degr$442
and other supergiants we have studied;
therefore, before making a comparison, we had to normalize the model and the observed profiles
at the pseudo-continuum regions between the neighboring lines. 

\begin{figure}
   \centering
     \includegraphics[width=9.0cm]{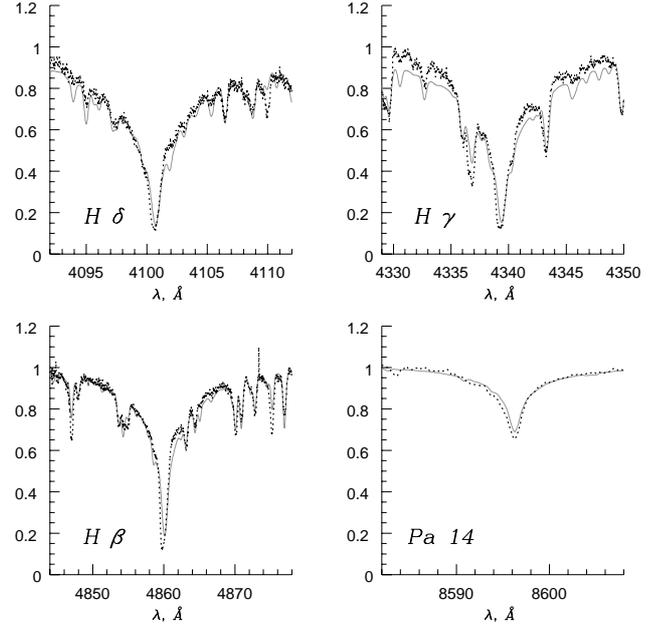}
      \caption{Hydrogen lines in BD+46$\degr$442 ({\it black dots})
compared to the best fitted of the available synthetic profiles ({\it grey solid lines}),
that were computed with $T_{\mathrm{eff}}=6250$ K, $\log g=2.0, \mathrm{[M/H]}=-0.5$.
To match the resolution of the model Paschen lines, the observed profiles were convolved to $R=20,000$, while
Balmer lines are shown at the native HERMES resolution of 85,000.
              }
         \label{fig_Balmer_model}
   \end{figure}

In the parameter space considered here,
model hydrogen lines strengthen both with higher $T_{\mathrm{eff}}$ and lower $\log g$ \citep{Fremat1996,Munari2000}.
Fortunately, a stronger sensitivity of 
Paschen lines to gravity allowed to exclude $\log g \ge 3.0$,
and  $T_{\mathrm{eff}} > 6500$ K along with it.
No Paschen models are available for $\log g<2$ for the range of temperatures
considered here, but they would require $T_{\mathrm{eff}} < 6000$ K,
which is too low to fit Balmer lines.
We therefore adopt $T_{\mathrm{eff}} = 6250 \pm 250$ K, $\log g < 3.0$ (Fig. \ref {fig_Balmer_model}), 
and return to the iron lines to put a better constraint on $\log g$.
Our estimate is further confirmed by a good match of Paschen lines and the shape
of H$\beta$ wings between BD+46$\degr$442 and Polaris (Fig. \ref{fig_Pol}).
Polaris Aa is a well-studied short-period Classical
Cepheid, for which e.g., \citet{Usenko2005} inferred
$T_{\mathrm{eff}} = 6000 \pm 170$ K, $\log g = 2.2 \pm 0.3$, and $\mathrm{[Fe/H]}=+0.07$.

%

\begin{figure}
   \centering
     \includegraphics[width=9.0cm]{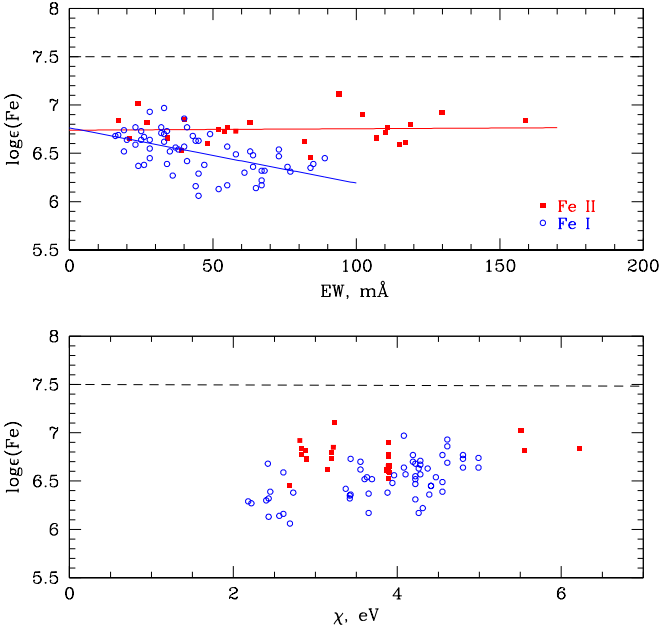}
      \caption{Dependence of iron abundance in BD+46$\degr$442 on EWs and the lower excitation potentials of individual lines
for our preferred model atmosphere $T_{\mathrm{eff}}/\log g/V_{\mathrm{tur}} = 6250/1.5/4.0$.
$T_{\mathrm{eff}}$ was constrained from fits to hydrogen lines,
$V_{\mathrm{tur}}$ from the requirement of no trend of individual \ion{Fe}{II} abundances
with EWs, and $\log g$ from the agreement between iron abundance inferred from \ion{Fe}{II} and weak \ion{Fe}{I} lines
(strong \ion{Fe}{I} lines being more prone to non-LTE effects).
The lower panel is less conclusive due to the non-uniform coverage of the excitation potentials by the \ion{Fe}{II} lines.
The dashed line indicates the iron abundance of the Sun.
              }
         \label{fig_Fe}
   \end{figure}

Finally, $T_{\mathrm{eff}}$ is also frequently derived from the requirement
that the iron abundance derived from the \ion{Fe}{I} lines does not depend on $\chi$.
Applying this criterion to BD+46$\degr$442, we obtain $T_{\mathrm{eff}} \approx$7500 K,
which is inconsistent with the hydrogen fits.
However, as we discussed above, \ion{Fe}{I} lines may not be reliable in our star due to non-LTE effects.
\ion{Fe}{II} lines, on the other hand, do not cover a large enough range of excitation potentials.
We thus restricted ourselves to $T_{\mathrm{eff}}=$6000, 6250, 6500 K
and, using the EW analyses of iron lines,
deduced the following combinations of $\log g/V_{\mathrm{tur}}$
for these temperatures: 1.0/4.0, 1.5/4.0, 2.0/4.0,
with precision of 0.5 dex in $\log g$ (set by the model step size),
and 1 km~s$^{\mathrm{-1}}$ for $V_{\mathrm{tur}}$.
Our final best model of 6250/1.5/4.0 is shown in Fig. \ref{fig_Fe}.
Linear fits to the \ion{Fe}{II} and \ion{Fe}{I} abundances plotted against EWs
(and extrapolated to EW$=$0 for \ion{Fe}{I}) give 6.74 $\pm$ 0.07 and 6.76 $\pm$ 0.06
for the iron abundance, respectively. The other two best-fit models
are used for the estimation of the model-dependent errors on the abundances.

%

%

\subsection{Abundances}

\begin{figure}
   \centering
     \includegraphics[bb=30 20 350 270, width=9.8cm, clip]{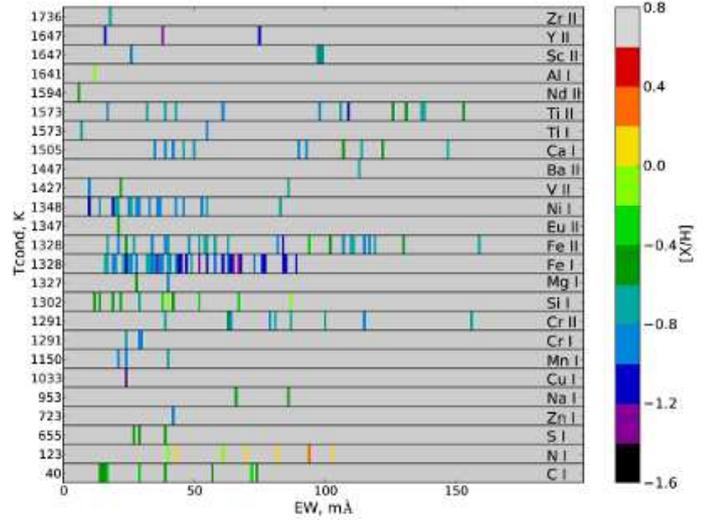}
      \caption{Chemical composition of BD+46$\degr$442 as inferred from the EW measurements
of individual lines in the 2009/08/17 spectrum
with the preferred model atmosphere $T_{\mathrm{eff}}/\log g/V_{\mathrm{tur}} = 6250/1.5/4.0$.
Colour-coded abundances are given relative to the Sun.
Elements are arranged in the order of increasing condensation temperature.
}
         \label{fig_abund_all}
   \end{figure}

\begin{figure}
   \centering
     \includegraphics[bb=20 160 500 550, width=11cm, clip]{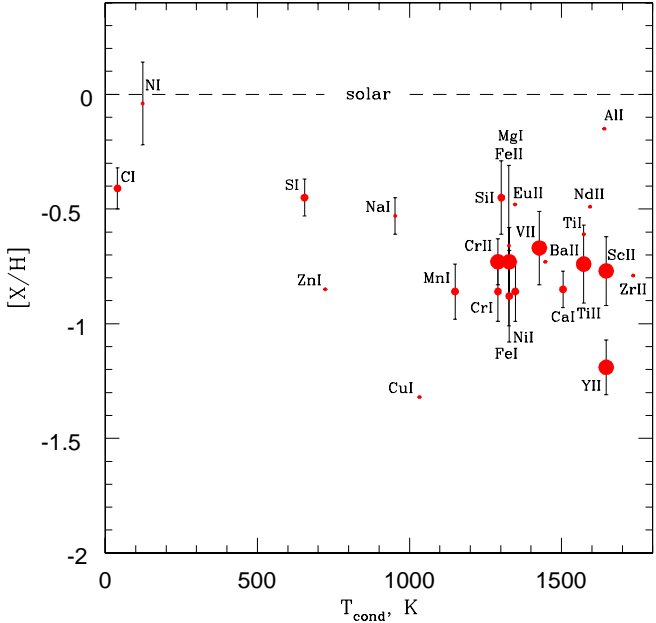}
      \caption{
Abundances of BD+46$\degr$442 averaged from several lines of a given element,
with error-bars representing the line-to-line RMS scatter.
The circle size designates the reliability of the lines against the non-LTE effects
(the bigger the more reliable, see text).
No significant depletion pattern is observed for this star;
week depletion is tentatively present, but better estimates are needed for the light elements
to confirm it. 
}
         \label{fig_abund_tcond}
   \end{figure}

The abundances of all elements deduced from the individual lines
using our best model atmosphere are shown in Fig. \ref{fig_abund_all}.
The elements are arranged in order of increasing condensation temperature
to verify the presence of the depletion pattern
reported in the literature for many disk sources.
The average abundance per element and the error budget are shown in Fig. \ref{fig_abund_tcond}
and in Table \ref{tab_errs}.
Given possible non-LTE effects, we assigned a quality flag
to each element (each ionization state), that reflects the reliability
of the final abundance estimate: flag 1 (best) -- ions that do not show a
trend with EW; flag 2 -- neutrals for which at least three lines with EW$\le$50 m\AA\,
were available for averaging; flag 3 (least reliable) -- all other cases.
These flags are reflected in the circle sizes in Fig. \ref{fig_abund_tcond}.
The errorbars are the RMS values,
they result from the uncertainties in the EW measurement and the values of $\log gf$-s.
Table \ref{tab_errs} also gives systematic errors due to the uncertainties in the model parameters.
Singly-ionized species are only weakly sensitive to variations in $T_{\mathrm{eff}}$ and $V_{\mathrm{tur}}$
compared to the neutral ones, but are more sensitive to $\log g$.

It is easy to see a moderate metal deficiency in BD+46$\degr$442,
at the level of $-0.5\ldots-0.7$ dex, compared to the Sun.
Apart from the slightly more abundant \rm{C} and \rm{S}, however,
there is no obvious trend with $T_{\mathrm{cond}}$.
The $\alpha$-process elements (\rm{Si, S, Mg, Ti}) show a 0.1--0.3 dex
enhancement over the iron group, while s-process elements (\rm{Y, Zr, Ba, Nd}) are not particularly enriched.
The high abundance of \rm{N} is likely a result of an evolutionary enrichment
during the first and the second dredge-ups, but can also be affected by the non-LTE effects \citep{Takeda1992,Przybilla2001}.
Overall this is a typical composition of a thick disk star without chemical depletion \citep{Reddy2006}.

%

%

%
\begin{table}
\caption{Chemical composition of BD+46$\degr$442.}             
\label{tab_errs}      
\centering          
\begin{tabular}{r c r r c c c c c}   
\hline     
Z     & Ion  & log $\epsilon$ & \rm{[X/H]} & RMS & \multicolumn{2}{c}{$\Delta \mathrm{[X/H]}$} & N & flag \\ 
\cline{6-7} 
\
        &      &                   &       &     & $A$ & $B$  & & \\ 
\hline 
6     & \ion{C}{I}  & 8.11 & $-$0.41 & 0.09 &$-$0.05 &$+$0.05    &6 & 2 \\
7     & \ion{N}{I}  & 7.89 & $-$0.04 & 0.18 &$+$0.02 &$-$0.01    &2 & 3 \\
11    & \ion{Na}{I} & 5.79 & $-$0.53 & 0.08 &$-$0.09 &$+$0.09    &2 & 3 \\
12    & \ion{Mg}{I} & 6.92 & $-$0.66 & 0.35 &$-$0.07 &$+$0.07    &2 & 3 \\
13    & \ion{Al}{I} & 6.32 & $-$0.15 & $-$  &$-$0.09 &$+$0.09    &1 & 3 \\
14    & \ion{Si}{I} & 7.09 & $-$0.45 & 0.16 &$-$0.08 &$+$0.07    &8 & 2 \\
16    & \ion{S}{I}  & 6.67 & $-$0.45 & 0.08 &$-$0.06 &$+$0.06    &3 & 2 \\ 
20    & \ion{Ca}{I} & 5.51 & $-$0.85 & 0.08 &$-$0.11 &$+$0.12    &5 & 2 \\
21    & \ion{Sc}{II}& 2.38 & $-$0.77 & 0.15 &$-$0.25 &$+$0.26    &4 & 1 \\ 
22    & \ion{Ti}{I} & 4.40 & $-$0.61 & $-$  &$-$0.17 &$+$0.16    &1 & 3 \\
22    & \ion{Ti}{II}& 4.27 & $-$0.74 & 0.17 &$-$0.25 &$+$0.24    &13& 1 \\
23    & \ion{V}{II} & 3.30 & $-$0.67 & 0.16 &$-$0.23 &$+$0.23    &3 & 1 \\
24    & \ion{Cr}{I} & 4.75 & $-$0.86 & 0.13 &$-$0.17 &$+$0.16    &3 & 2 \\ 
24    & \ion{Cr}{II}& 4.86 & $-$0.73 & 0.10 &$-$0.17 &$+$0.18    &9 & 1 \\ 
25    & \ion{Mn}{I} & 4.52 & $-$0.86 & 0.12 &$-$0.14 &$+$0.14    &3 & 2 \\ 
26    & \ion{Fe}{I} & 6.59 & $-$0.88 & 0.20 &$-$0.13 &$+$0.13    &39& 2 \\
26    & \ion{Fe}{II}& 6.75 & $-$0.73 & 0.15 &$-$0.18 &$+$0.19    &25& 1 \\ 
28    & \ion{Ni}{I} & 5.37 & $-$0.86 & 0.13 &$-$0.14 &$+$0.13    &15& 2 \\  
29    & \ion{Cu}{I} & 2.89 & $-$1.32 & $-$  &$-$0.20 &$+$0.21    &1 & 3 \\ 
30    & \ion{Zn}{I} & 3.72 & $-$0.85 & $-$  &$-$0.16 &$+$0.18    &1 & 3 \\ 
39    & \ion{Y}{II} & 1.03 & $-$1.19 & 0.12 &$-$0.27 &$+$0.27    &3 & 1 \\ 
40    & \ion{Zr}{II}& 1.79 & $-$0.79 & $-$  &$-$0.25 &$+$0.25    &1 & 3 \\ 
56    & \ion{Ba}{II}& 1.40 & $-$0.73 & $-$  &$-$0.31 &$+$0.30    &1 & 3 \\ 
60    & \ion{Nd}{II}& 1.07 & $-$0.49 & $-$  &$-$0.28 &$+$0.29    &1 & 3 \\ 
63    & \ion{Eu}{II}& 0.02 & $-$0.48 & $-$  &$-$0.26 &$+$0.27    &1 & 3 \\  
\hline                  
\end{tabular}
\tablefoot{Abundances are given for our best photospheric model
$T_{\mathrm{eff}}/\log g/V_{\mathrm{tur}} = 6250/1.5/4.0$. $\Delta \mathrm{[X/H]}$ designates abundance changes
corresponding to the changes in T$_{\mathrm{eff}}$/$\log g$ of $-$250 K/$-$0.5 dex (case A)
and $+$250 K/$+$0.5 dex (case B), which are our second best atmospheric model estimates.
The flag designates the quality of the N lines used in the averaging:
from 1 for the best cases to 3 for the worst. Elements with flag 3 had less
than 3 lines with EW$\le$50 m\AA\,. }

\end{table}

\section{Spectral energy distribution}\label{sec_sed}

\begin{figure}
   \centering
      \includegraphics[bb=25 407 400 720, width=9.6cm, clip]{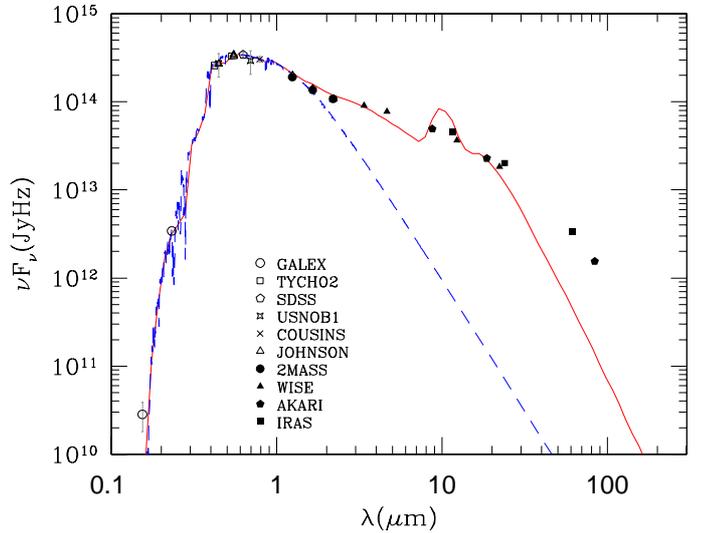}
      \caption{
The observed spectral energy distribution of BD+46$\degr$442
plotted against the reddened Kurucz's model
({\it dashed line)} and our star$+$disk model ({\it solid line}).
The flux up to 30 $\mu$m can be reproduced with $10^{-6}\, M_{\odot}$
of dust in grains smaller than 100 $\mu$m in size that reside in a ring
that stretches from 7 AU to $\sim$500 AU.
}\label{fig_SED}
   \end{figure}

Fig. \ref{fig_SED} compares photometric observations of BD+46$\degr$442
to the reddened photospheric Kurucz model
with stellar parameters as derived in Sec. \ref{sec_phot}.
The total reddening of $E(B-V)=0.18$ was obtained by minimizing the difference between the adopted reddened
model and the observed optical and near-infrared fluxes.
The presence of the excess emission beyond 2 $\mu$m is obvious.
When we selected BD+46$\degr$442 for our sample,
we could only rely on the IRAS detection of the excess (of which the 100 $\mu$m flux is still an upper limit),
and now it is confirmed  with the two other space missions, AKARI \citep{Ishihara2010} and WISE \citep{Wright2010}.
The SED, which is clearly not double-peaked and only slightly reddened,
argues against a simple detached shell distribution for the circumstellar material.
Instead, the SED of this kind is characteristic of a star surrounded by a passive disk \citep[e.g, in][]{Deruyter2006, Gielen2007}.

We modeled the SED of BD+46$\degr$442 with MCMax \citep{Min2009}, a 2D Monte Carlo radiative transfer disk code.
This code computes the temperature structure and density of the disk.
The vertical scale height of the disk is computed by an iteration process,
demanding vertical hydrostatic equilibrium.
For the input stellar model we use the parameters as discussed in Section \ref{sec_abunds}
and adopt a total system mass $M_{\mathrm{\star}}=2\, M_{\odot}$ and a typical post-AGB
luminosity of $L_{\mathrm{\star}}=2000\, L_{\odot}$ (corresponding to a distance for the system of 3.3 kpc).
Since we lack information on the exact mineralogy of the dust in the disk, we use
a mixture of astronomical silicates and metallic iron, with grain sizes ranging from 0.1 to 100 $\mu$m.
The adopted gas-to-dust ratio is kept fixed at 100.
The geometry of the disk is determined by the inner and outer radii $R_{\mathrm{in}}$ and $R_{\mathrm{out}}$,
the total dust mass $M_{\mathrm{dust}}$, and the power law of the surface-density distribution of the dust
$\Sigma \sim R^{-p}$.

To keep the numbers of free parameters for the model low, we keep the outer radius fixed
at $R_{\mathrm{out}}=500\,$AU, and use a value $0.5 < p < 2.0$ for the surface-density distribution,
as expected in a disk environment. To account for the low observed total reddening,
the inclination is kept fixed at $15\degr$.
This modeling is quite degenerate and equally well fitting models with slightly
different sizes, total masses and surface-density distributions can be found.
Complementary measurements, such as interferometric observations and infrared spectroscopy,
would be invaluable to constrain further the disk geometry and mineralogy.

In Fig. \ref{fig_SED} we show a disk model which gives good agreement with the photometric data.
The disk parameters for this model are: $R_{\mathrm{in}}=7$\,AU, $R_{\mathrm{out}}=500\,$AU, $p=1.6$, and $M_{\mathrm{dust}}=2 \times 10^{-6}\, M_{\odot}$.
For the dust composition we use a combination of amorphous olivine and metallic iron,
in a ratio of $\sim 96\% / 4\%$.
The AKARI and IRAS fluxes past 30 $\mu$m can not be fit by this model
because that part of the SED requires even larger dust grains than adopted.
These large grains are likely not uniformly mixed in the disk,
but must have settled towards the mid-plane.
Modeling such process is beyond the scope of our investigation. 

BD+46$\degr$442 thus belongs to a distinct group of evolved stars
surrounded by dusty disks \citep{Gielen2011}. To verify whether it is
a binary system, and whether the disk is circumstellar
or circumbinary, we analyze the radial velocities.

%

\section{RV curve}\label{sec_rv}

\begin{figure}
   \centering
    \includegraphics[width=10cm]{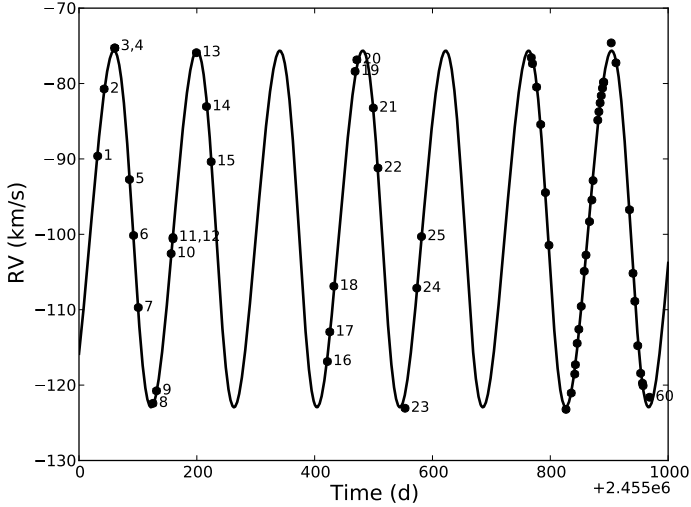}
      \caption{
HERMES radial velocities of BD+46$\degr$442 plotted against Keplerian orbit
with parameters listed in Table \ref{tab_RVparms}.
}\label{fig_RVs}
   \end{figure}

We measured RVs using a cross-correlation
method with a mask representing a G2 star.
The mask consists of $\sim$1100 lines.
Cross-correlation functions (CCFs) extracted from each order were averaged,
and the resultant profile
was fitted with a Gaussian to determine the RV.
Some asymmetries in the CCF were noted that
appear to correlate with the orbital phase (see discussion
in the following section). They, however, were
never large enough to affect the location of the main peak.
The measured velocities are given in Table \ref{tab_RVs} and plotted in Fig. \ref{fig_RVs}.
Errors in the RV consist of two quadratically added terms: the error of the Gaussian fit
and the 0.2 km~s$^{\mathrm{-1}}$ uncertainty due to the drift of the wavelength calibration zero-point.

%

As one can see from Fig. \ref{fig_RVs}, the three seasons of observations reveal
a regular variation in the RV with a period $\sim$100 days and
peak-to-peak amplitude of $\sim$40 km~s$^{\mathrm{-1}}$.
Integrating over half a period under the RV curve, we obtain a displacement of 130 $R_{\odot}$,
which is more than the radius of a post-AGB star. The radial velocity
variations cannot be due to pulsations.

\begin{figure}
   \centering
     \includegraphics[width=10cm]{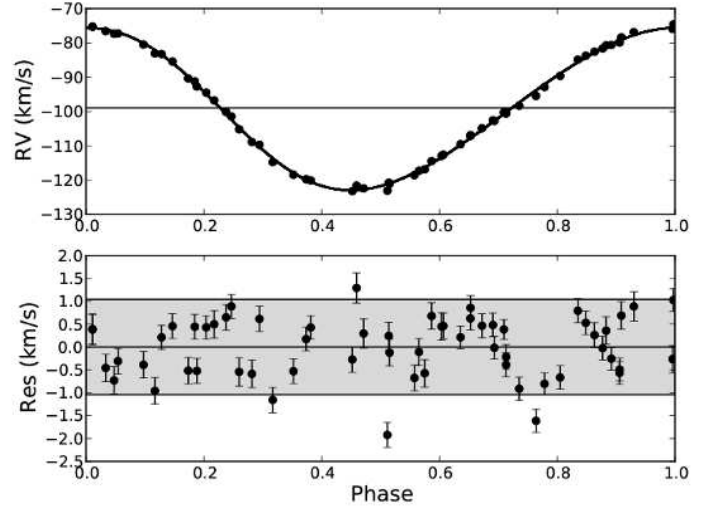}
      \caption{
Upper panel: radial velocities of BD+46$\degr$442 phased with a period of 140.77 days
and overlaid on our Keplerian fit. The horizontal line marks the corresponding
systemic velocity.
The $\phi=0$ was defined to correspond to the maximum RV (i.e. near one of the elongations).
Bottom panel: residuals from the fit with $\pm 3\sigma$ margins.
}
         \label{fig_RVPhase}
   \end{figure}

\begin{figure}
   \centering
    \includegraphics[width=9.5cm]{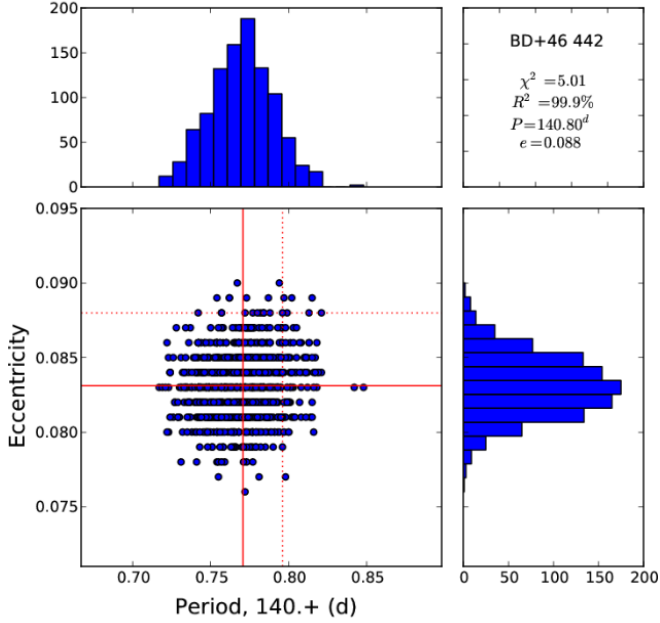}
    \caption{
Distribution of the orbital period and eccentricity values
obtained from 1000 simulated RV data sets. The solid cross marks the solution based on the
observed RVs (see Table \ref{tab_RVparms}),
while the dotted cross and the upper right panel show an alternative solution
with a slightly better $\chi^{2}$ derived from the simulated RV data sets.
}
    \label{fig_eTee}
   \end{figure}

A long period with a nearly-sinusoidal shape of the RV curve, near-constant temperature,
and the large displacement of the photosphere
speak strongly against pulsations.
Rather, we must be observing an orbital motion around an invisible companion.
BD+46$\degr$442 is thus an SB1 system.
Indeed, we obtained a very good Keplerian solution
with only one point (spectrum 23) out of 60 deviating by more than 3$\sigma$
(Fig. \ref{fig_RVPhase}).
The derived orbital parameters are given in Table \ref{tab_RVparms}.
The errors on the parameters were estimated via the Monte Carlo method,
using 1000 simulated RV sets. For a given date a value for the RV was drawn from
a Gaussian distribution with a mean
and $\sigma$ equal to the observed RV value and its error, respectively.
For every simulated dataset the orbital parameters were obtained,
and their scatter was used as an error estimate
on the parameters obtained from the real data.
An example application of this procedure for the orbital period
and eccentricity is given in Fig. \ref{fig_eTee}.
As one can see, albeit small (0.08), the eccentricity is significantly different from zero.

Our observations provide the first account as far as we know
of the RV variations in BD+46$\degr$442, that we successfully reproduce
with a Keplerian orbit. BD+46$\degr$442 proves to be yet another evolved system with a disk that is a binary.
The semi-major axis of the primary is at most 1.2 AU
(for $i=15\degr$). This is smaller than the silicate dust sublimation radius
that would be at 2 AU (given $L_{\mathrm{\star}}=2000\, L_{\odot}$), and much smaller than the actual
inner disk radius at 7 AU estimated from the fit to the NIR excess with metallic iron particles.
This barely leaves space for a cicumstellar dusty disk around any of the binary components,
unless the gas is very optically thick to shield some radiation.
We conclude that the dusty disk producing the IR excess is circumbinary rather than circumstellar.

%

%
\onltab{1}{
\begin{table}
\caption{Radial velocities of BD+46$\degr$442}             
\label{tab_RVs}      
\begin{tabular}{r r r r}   
\hline     
N    & JD 2,455,000$+$ & RV     & $\sigma$  \\ 
     & (d)             & (km~s$^{\mathrm{-1}}$) & (km~s$^{\mathrm{-1}}$)    \\ 
\hline 
 1 &  31.71 &  -89.62 & 0.26  \\
 2 &  42.70 &  -80.72 & 0.30  \\
 3 &  60.70 &  -75.28 & 0.33  \\
 4 &  60.72 &  -75.29 & 0.33  \\
 5 &  85.64 &  -92.73 & 0.27  \\
 6 &  92.64 & -100.15 & 0.28  \\
 7 & 100.62 & -109.71 & 0.28  \\
 8 & 125.49 & -122.42 & 0.32  \\
 9 & 131.53 & -120.76 & 0.30  \\
10 & 156.40 & -102.56 & 0.26  \\
11 & 159.55 & -100.60 & 0.26  \\
12 & 159.56 & -100.42 & 0.27  \\
13 & 199.33 &  -75.91 & 0.30  \\
14 & 216.38 &  -83.05 & 0.29  \\
15 & 224.36 &  -90.37 & 0.29  \\
16 & 421.63 & -116.87 & 0.30  \\
17 & 425.67 & -112.94 & 0.30  \\
18 & 432.59 & -106.89 & 0.27  \\
19 & 468.66 &  -78.39 & 0.30  \\
20 & 471.62 &  -76.86 & 0.32  \\
21 & 499.54 &  -83.24 & 0.26  \\
22 & 507.47 &  -91.20 & 0.27  \\
23 & 553.48 & -123.07 & 0.27  \\
24 & 573.34 & -107.14 & 0.25  \\
25 & 581.34 & -100.30 & 0.22  \\
26 & 767.71 &  -76.56 & 0.30  \\
27 & 769.71 &  -77.36 & 0.30  \\
28 & 776.73 &  -80.47 & 0.29  \\
29 & 783.69 &  -85.43 & 0.27  \\
30 & 791.74 &  -94.47 & 0.25  \\
31 & 797.73 & -101.45 & 0.26  \\
32 & 826.65 & -123.22 & 0.28  \\
33 & 835.57 & -121.05 & 0.29  \\
34 & 841.50 & -118.54 & 0.28  \\
35 & 842.59 & -117.29 & 0.30  \\
36 & 845.61 & -114.46 & 0.29  \\
37 & 848.41 & -112.60 & 0.28  \\
38 & 852.52 & -109.55 & 0.25  \\
39 & 857.64 & -104.90 & 0.26  \\
40 & 860.60 & -102.75 & 0.25  \\
41 & 866.54 &  -98.31 & 0.25  \\
42 & 870.56 &  -95.45 & 0.25  \\
43 & 872.61 &  -92.87 & 0.24  \\
44 & 880.56 &  -84.86 & 0.26  \\
45 & 882.46 &  -83.74 & 0.26  \\
46 & 884.55 &  -82.57 & 0.27  \\
47 & 886.49 &  -81.61 & 0.26  \\
48 & 888.55 &  -80.62 & 0.25  \\
49 & 890.52 &  -79.87 & 0.25  \\
50 & 890.54 &  -79.79 & 0.25  \\
51 & 903.38 &  -74.61 & 0.25  \\
52 & 911.41 &  -77.25 & 0.28  \\
53 & 934.41 &  -96.74 & 0.30  \\
54 & 940.36 & -105.17 & 0.31  \\
55 & 943.44 & -108.88 & 0.30  \\
56 & 948.40 & -114.77 & 0.28  \\
57 & 953.36 & -118.43 & 0.27  \\
58 & 956.40 & -119.74 & 0.26  \\
59 & 957.48 & -120.08 & 0.26  \\
60 & 968.45 & -121.61 & 0.33  \\
\hline                  
\end{tabular}
\tablefoot{Radial velocities (barycentric correction included)
measured with HERMES over the period July 2009 -- February 2012.
} 
\end{table}
}

%


\begin{table}
\caption{Orbital elements of BD+46$\degr$442}             
\label{tab_RVparms}      
\centering          
\begin{tabular}{l r r }   
\hline     
Parameter  & Value  &  $\sigma$  \\ 
     &  &            \\ 
\hline 
$P$ (d)            & 140.77      & 0.02 \\
$a$ $sini$ (AU)    & 0.31        & 0.001 \\
$f(m)$ ($M_{\odot}$) &  0.19       & 0.001 \\
$K$ (km~s$^{\mathrm{-1}}$)         & 23.66       & 0.06 \\
$e$                & 0.08        & 0.002 \\
$\omega$ ($\degr$) & 100         & 2 \\ 
$T_{\mathrm 0}$ (JD)        & 2 455 094.6 & 0.6 \\
$\gamma$ (km~s$^{\mathrm{-1}}$)    & -98.96      & 0.04 \\
$\chi^{2}$  & 6.4 & \\
$R^{2}$ & 99.83\% & \\
\hline                  
\end{tabular}
\tablefoot{Orbital parameters with their uncertainties
and two statistical parameters indicating the goodness of the RV fit
(reduced chi-squared and the coefficient of determination).
} 
\end{table}

\section{Line profiles}\label{profs}

   \begin{figure*}[htpb]
   \centering
     \includegraphics[width=9cm]{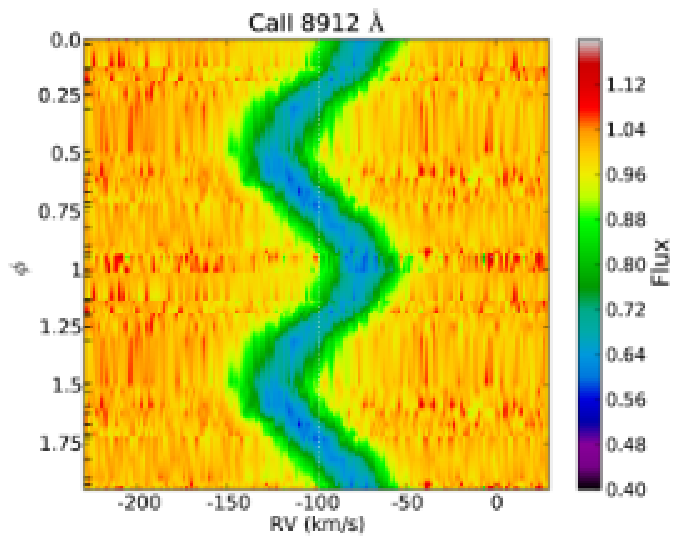}
     \includegraphics[width=9cm]{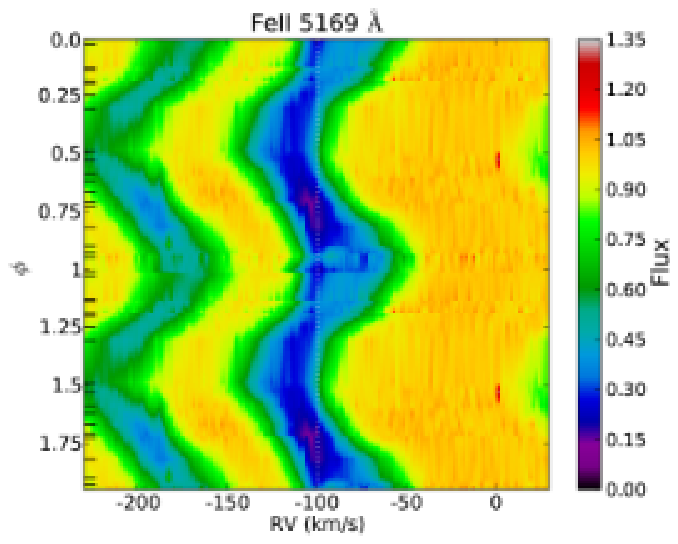}
     \includegraphics[width=9cm]{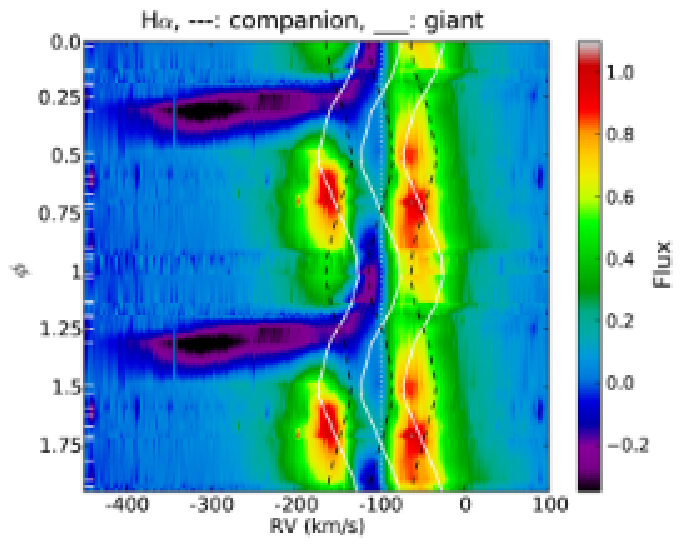}
     \includegraphics[width=9cm]{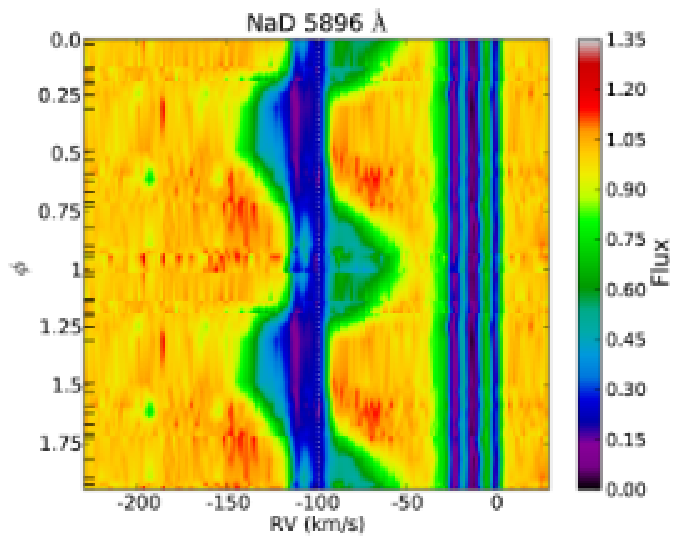}
     \includegraphics[width=9cm]{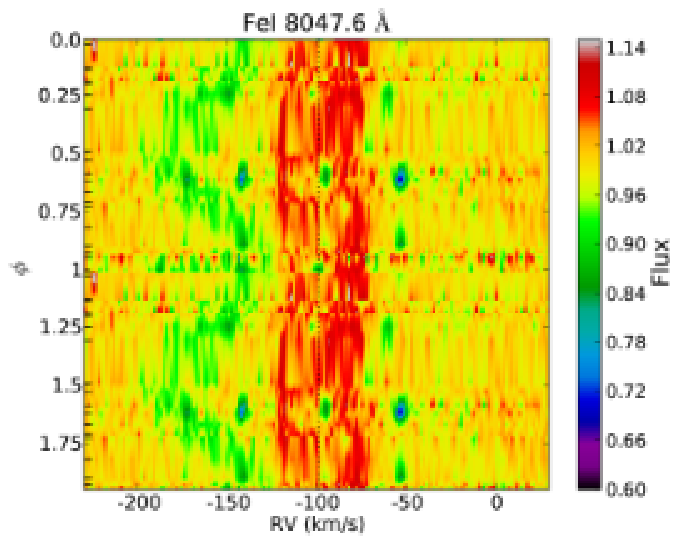}
     \includegraphics[width=9cm]{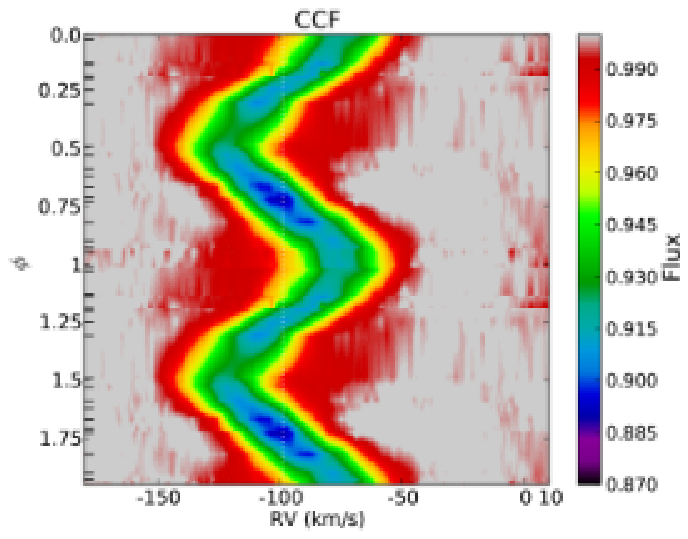}
      \caption{Dynamic spectra of selected lines and the CCF as a function of the orbital phase (time runs
from top to bottom).
Colours designate continuum-normalized fluxes (except for H$\alpha$), with black corresponding to the strongest absorption.
Fluxes in the missing phases were obtained by a linear interpolation between the nearest
observed phases. Dotted vertical line marks the systemic velocity,
while the horizontal dashes on the left -- the observed phases. One orbital period is shown twice to guide the eye.
H$\alpha$ is plotted after subtraction of the photospheric model spectrum; solid and dashed lines represent
the RV curves (original and $\pm$50 km~s$^{\mathrm{-1}}$ offset) of the giant and of the putative companion to illustrate
the lack of significant motion in the emission component. Only data from the first two seasons of observations
are shown (spectra 1-25, obtained between July 2009 - January 2011); later observations confirm the general
behaviour with phase, but the intensities of the circumstellar features appear to vary slightly from cycle to cycle. 
              }
         \label{fig_all2d}
   \end{figure*}
%

\subsection{H$\alpha$}

In Fig.\ref{fig_Balmer} we show that H$\alpha$ 
alternates between two major profiles:
a double-peak emission with peaks separated by $\sim$100 km~s$^{\mathrm{-1}}$
and wings stretching to $\pm$200 km~s$^{\mathrm{-1}}$, and a P~Cyg-like profile.
This is better illustrated in Fig.\ref{fig_all2d} (medium left panel) where
we depict H$\alpha$ after subtraction of
the model photospheric profile. The resulting profile shows
both emission and absorption components
that vary in strength with the orbital phase.
The double-peak emission is observed most of the time,
but is strongest and most symmetric between $\phi = 0.7 - 0.8$,
corresponding to the giant's inferior conjunction (when it is between us and the companion, Fig. \ref{fig_cart}).
When the giant starts to retreat, the emission diminishes,
while the central absorption
blue-shifts and broadens, quickly turning
into a P~Cyg profile. The expansion velocities reach 350 km~s$^{\mathrm{-1}}$ at $\phi = 0.3 - 0.4$
when the companion is in front.
Higher members of the Balmer series develop broadened cores at this phase,
but the effect is much less pronounced than in H$\alpha$, 
particularly in the wings (Fig. \ref{fig_Balmer}), which justifies their use
for estimating T$_{\mathrm{eff}}/\log g$ .
The H$\alpha$ profile as a whole does not seem to shift in RV, but
is not centered on the systemic velocity ($V_{\mathrm{syst}}$) either,
rather, it is permanently blue-shifted by $\sim$14 km~s$^{\mathrm{-1}}$.

\subsection{Non-photospheric components of metal lines}

In the other panels of Fig.\ref{fig_all2d} (and in Fig. \ref{fig_multiplets})
we show the CCF and a representative set of metal line profiles.
As expected, the CCF follows the behaviour
of weak to medium-strong lines, like \ion{Ca}{II} 8912 \AA, that constitute the majority of unblended lines
in the CCF list. These lines are ``well behaved'', being symmetric and hardly
variable, except for the RV that reflects the orbital motion of the giant.
The \ion{Fe}{II}(42) triplet, on the other hand, presents an example of strong lines
(EW$\gtrsim$300 m\AA) that show an additional narrow component near $V_{\mathrm{syst}}$.
It is best seen when the giant retreats from us ($\phi =0.9 - 1.1$),
while in the opposite phase it appears less prominent, being either weaker or blue-shifted
(by $\sim$10 km~s$^{\mathrm{-1}}$ from $V_{\mathrm{syst}}$) and overlapping with the photospheric component.
Other similar cases involve strong lines of \ion{Ba}{II}, \ion{Ca}{II},
\ion{Fe}{I}, \ion{Fe}{II}, \ion{Mg}{I}, and \ion{Ti}{II}, and the resonance lines of  \ion{K}{I}.
Exceptions include the prominent \ion{O}{I} triplets at 7775 and 8446 \AA\,,
and the \ion{Ca}{II} doublet at 8915 \AA\,, that still show normal, single-deep, profiles.
These transitions, however, require much higher excitation ($\chi_{\mathrm{low}}=$7 -- 9 eV)
and form deeper in the atmosphere.
The strengthening of the central absorption between $\phi=0.8-1.25$
and the disappearance or blue-shift between $\phi=0.3-0.7$ resemble
the behaviour of the central absorption in H$\alpha$.
H$\alpha$ absorption, however, is more blue-shifted, particularly in the P~Cyg phase.

\ion{Na}{I} D lines also show a central absorption component, but also an additional,
narrower one 13 km~s$^{\mathrm{-1}}$ to the blue from $V_{\mathrm{syst}}$, which coincides with the centroid of H$\alpha$.
Both components are probably of circumstellar origin, but it is difficult
to study them due to blending with the photospheric lines.
To the red, there are three supposedly ISM components,
at RV$=-$23, $-$13, and $-$1 km~s$^{\mathrm{-1}}$, that are stable in RV and depth.
They are also observed in the resonance \ion{K}{I} lines. 
Using a code provided in \citet{Reid09}, we calculated kinematic distances to the
foreground clouds responsible for these components. The largest distance is obtained for the
most blue-shifted component ($V_{\mathrm{helio}}=-23$ km~s$^{\mathrm{-1}}$, $V^{rev}_{\mathrm{LSR}}=-18$ km~s$^{\mathrm{-1}}$),
which is 1.25 $\pm$ 0.5 kpc. Taking this as a minimum distance to BD+46$\degr$442 itself,
and correcting for the reddening $A_{V}=0.56$, we obtain a lower limit on the stellar luminosity:
$M_{\mathrm{V}} = -1.54^{+1.19}_{\,-0.81}$. Clearly, the star is brighter than normal FG giants
of the luminosity class III ($M_{\mathrm{V}} = 1\ldots2$), but is consistent with being an F type post-AGB star,
which have $M_{\mathrm{V}} = -2\ldots-3$ \citep{Bond1996, Ginestet02}.
On the other hand, the two components at $V_{\mathrm{helio}} \sim -100$ km~s$^{\mathrm{-1}}$
would indicate an Iab supergiant $(M_{\mathrm{V}}= -6\ldots-7)$ at more than 2.5 kpc above the Galactic plane,
which is very unlikely.
Thus, the two components near $V_{\mathrm{syst}}$ are most likely circumstellar.

In addition to the absorption components, as can be seen from Fig. \ref{fig_all2d},
near $\phi=$0.75 corresponding to the giant's inferior conjunction,
weak emission appears in the wings of the sodium lines, echoing H$\alpha$.
We observe only one other system of emission lines beside H$\alpha$ and \ion{Na}{I} D:
the weak \ion{Fe}{I}(12) multiplet at 8047.6 and 8075.1 \AA.
Similar to the central absorption of metal lines, the emission is centered on the systemic velocity
and does not participate in the orbital motion (Fig. \ref{fig_all2d}).
 
\subsection{Secondary spectrum ?}

The metal lines noticeably strengthen near the giant's inferior conjunction
($\phi = 0.75$). 
Zooming in on the wings of the CCF (see the right-bottom panel of Fig. \ref{fig_all2d}), however, reveals
also faint absorption spectrum moving in anti-phase with the giant's spectrum.
Therefore, we may be observing a companion spectrum instead.
We remodeled the CCF profiles with two Gaussians, where possible, and obtained
a semi-amplitude of 15$\pm$5 km~s$^{\mathrm{-1}}$ for the secondary spectrum.
Dividing by the 23.7 km~s$^{\mathrm{-1}}$ of the primary, we obtain
a mass ratio of 0.6. Given this mass ratio and the mass function of 0.2,
we obtain the individual masses of 0.9--0.3 $M_{\odot}$ for the giant
and 1.4--0.5 $M_{\odot}$ for the companion
for a range of inclinations between 45--90$\degr$.
These masses are consistent with a low-mass pair where
the originally more massive component has evolved into a post-AGB,
while the companion is still on the MS or in the sub-giant stage.
Using expressions for the Roche-lobe radius from \citet{Paczynski1971},
that depend only on the mass ratio, we obtain for the giant
$R_{1}\approx0.37\times(a_{1}+a_{2})=57-39 R{_{\odot}}$.
The radius of BD+46$\degr$442 matches this range well if it has a typical post-AGB
luminosity of 2000-3000 $L_{\odot}$. Therefore, we are observing an interacting system
where the post-AGB star is (close to) filling its Roch-lobe and as
such feeding a more compact companion.

We want to stress, however, that these are very preliminary estimates.
The depth of the secondary component in the CCF is about 1\%
(compared to 10\% of the giant's spectrum),
which is far below the typical S/N ratio of our spectra.
The presence of this component thus can not be verified from individual lines.
Furthermore, the spectral type of the companion
can not be significantly different from that of the primary,
otherwise one would not detect it with the G2 mask.
The companion then should be only ten times fainter than the giant primary,
which is too bright were it a MS star.
And if both components are giants,
they would have to have nearly identical initial masses
due to the short longevity of the giant stage.
It is possible instead that we are seeing a pseudo-photosphere of
an accretion disk around the companion rather than the companion's spectrum itself.
Higher S/N spectra and a multi-band photometry
are needed to confirm the contribution of the companion in the system's flux.

\section{Discussion}

The discovery of the RV variations in BD+46$\degr$442 provides a strong support
for the disk-binary connection in post-AGB stars.
With a period of $140.77 \pm 0.02\, ^{\rm d}$ it is one of the shortest known post-AGB binaries.
A small but significant eccentricity indicates that the system is not completely circularized.
In the following discussion we will show how the observed spectral features
in BD+46$\degr$442 indicate an ongoing mass transfer
from the giant to the companion via the Roche-lobe overflow.

\subsection{H$\alpha$ variability in interacting binaries}

The H$\alpha$ profiles discussed here are not uncommon among post-AGB stars
\citep{Waters1993a,Pollard1997,Maas2005,Sanchez2008}, but they have not been systematically
studied for variability in function of orbital phase. We are aware of four other disk systems
where a strong blue-shifted absorption was noted to develop during the giant's superior conjunction:
HR 4049 with $P_{\mathrm{orb}}=430^{\rm d}$, $a\sin i=0.6$ AU, $i \ge 60\degr$, $e=0.3$ \citep{Waelkens1991, VanWinckel1995, Bakker1998};
HD 44179 (the central star of the Red Rectangle nebula) with $P_{\mathrm{orb}}=318^{\rm d}$, $a\sin i=0.5$ AU, $i_{\mathrm{"effective"}} = 35\degr$,
$e=0.4$ \citep{Waelkens1996, Witt2009};
IRAS 08544$-$4431 with $P_{\mathrm{orb}}=508^{\rm d}$, $a\sin i=0.4$ AU, $i \sim 60\degr$, $e=0.2$
\citep{Maas2003, Deroo2007, VanWinckel2009}; IRAS 19135$+$3937 with $P_{\mathrm{orb}}=127^{\rm d}$, $a\sin i=0.2$ AU,
$e\sim0.3$ \citep{Gorlova2011}.
The fact that BD+46$\degr$442 has a very small eccentricity ($e = 0.083 \pm 0.002$)
indicates that the cause of the spectral variations most likely lies
in the varying line of sight towards the components,
rather than in the physical changes during the orbital motion,
such as an increased mass loss at periastron.
 
The most common interpretations of the double-peak H$\alpha$ emission
in stellar spectra are:
an inclined Keplerian gaseous disk (e.g. in Be and T Tau stars),
the giant's atmosphere that is irradiated by a hot companion (Cataclysmic Variables, Symbiotics),
propagating shock in pulsating stars (Miras, RV Tau stars), and the chromosphere.
There is no indication in our spectra of a hot companion
(no helium lines or nebular emission),
pulsations are excluded based on the smooth orbital RV curve,
and the existence of chromospheric activity has yet to be
established in post-AGB stars.
We therefore concentrate on the disk hypothesis.
The emission could originate in the gaseous extension of the circumbinary
dusty disk towards the center of the system or in a circumstellar
disk around one of the components.
To differentiate between these possibilities
we searched for similarities with other types of interacting binaries.
There, a disk sometimes forms around a more compact companion as a result of
the Roche-lobe overflow or a wind accretion from a larger companion.

In Algols, when the binary separation is large, the accretion stream
from a cool giant curves and settles into a disk around
a more massive hot MS primary \citep{Richards99}.
The simulations of \citet{Miller2007} show that the dominant source
of H$\alpha$ emission in Algols is the disk, while the
stream provides a much smaller contribution.
A similar situation could be occurring in BD+46$\degr$442,
only the accreting companion is much fainter than the giant (Fig. \ref{fig_cart}).
The problem with this interpretation of the H$\alpha$ emission
is the lack of convincing motion in the anti-phase with
the absorption spectrum of the primary (Fig. \ref{fig_all2d}).
The P~Cyg-like profile, on the other hand, is a signature of an outflow.
It is very rare to find such broad absorption feature (with velocities
up to -300 km~s$^{\mathrm{-1}}$) in Algols, and it may not be observed
in every cycle \citep{Peters1989, Miller2007}.
Usually Algols display classical and inverse P~Cyg-like profiles at a smaller range of velocities
just before and after giant's inferior conjunction,
which is explained by the disk eclipse.
This is clearly not the case in BD+46$\degr$442
where the wind profile is strongest near the giant's superior conjunction.
The explanation of the P~Cyg profile, therefore, should be searched elsewhere.

W Ser systems are thought to be in the stage preceding Algols,
when the transfer rate is very high and the accretion disk completely obscures the accretor.
In at least one such system, $\upsilon$ Sgr with  $P_{\mathrm{orb}}=137^{\rm d}$ \citep{Netolicky2009},
a P~Cyg-like profile is observed near the giant's superior conjunction,
which \citet{Nariai1967} explained by a ``coronal stream'' from the giant
to the companion. At this orbital phase the stream
is directed towards us and at the same time is projected against the giant (Fig. \ref{fig_cart}),
creating a blue-shifted absorption.
Furthermore, in the simulations of interacting binaries the accretion stream
is often found to be curved in the direction of the giant's motion,
which could explain why the largest speeds in BD+46$\degr$442
are reached at a slightly later phase after the conjunction
(compare spectra 6 and 7 on Fig. \ref{fig_Balmer}).
Interestingly, $\upsilon$ Sgr is considered to be a non-eclipsing type of W Ser stars.
In classical, edge-on, systems the blue-shifted absorption is not observed,
perhaps because the accretion stream, when projected against the donor,
is hidden from our view by the accretion disk.

Symbiotic stars are another relevant class of objects to consider.
There double-peaked emission profiles are quite common \citep{VanWinckel1993},
and are being interpreted as either due to an accretion disk around a WD
\citep{Robinson1994, VanEck2002} or a disk wind \citep{Skopal2006}.
There are even reports that the blue peak gets suppressed 
at the orbital phase corresponding to the occurrence of
the P~Cyg-like profile in BD+46$\degr$442 \citep{Robinson1994, Murset2000},
but the true broad absorption in symbiotics is only observed in outbursts.

\subsection{Jet launching in the accretion disk of the companion}

Summarizing, the variations of H$\alpha$ in BD+46$\degr$442 are complex, but can be
decomposed into several time-dependent properties for which we need an explanation.
1) The double-peak emission is rather stable in velocity along the orbit,
which means that it is connected to the system itself and not to one
of the components. The emission, however, varies in strength,
being strongest during the giant's inferior conjunction.
2) The P~Cyg-like absorption component is observed at very high
velocity, but only during a very short orbital phase interval, when the giant is at
the superior conjunction (see Fig.\ref{fig_all2d}).
As we have seen in the previous section, some of these aspects
can be explained by the interaction with the companion.

\begin{figure}[hbtp]
   \centering
    \includegraphics[bb=40 116 310 420, scale=1.0, clip]{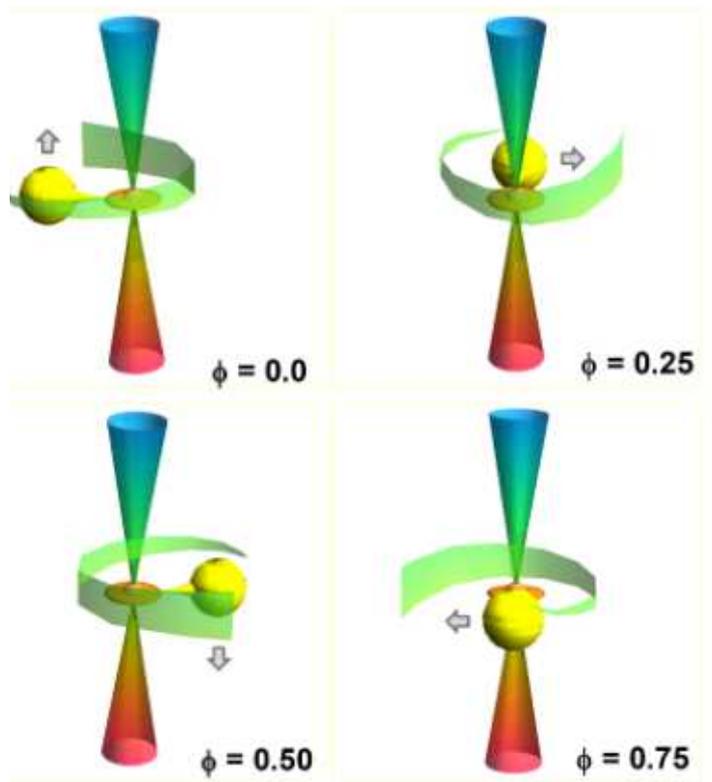}
    \caption{
Schematic representation (not to scale) of the views of BD+46$\degr$442 at different orbital phases.
The dusty circumbinary disk has been omitted
to better illustrate gas flows in the vicinity of the components (the inner disk wall
would be outside the boundaries of the figure anyway).
The inclination, both of the orbit and of the likely precessing jet, remains to be constrained.
The arrow shows the direction of the giant's motion.
The figure illustrates how an accretion jet around the companion
can explain the transient P~Cyg-like profile of H$\alpha$,
while a gas stream trailing behind the giant can explain the narrow absorption component of metal lines.
}
    \label{fig_cart}
   \end{figure}

It is useful to compare this behaviour to the line variability as
observed and discussed in HD 44179 by \citet{Witt2009}. 
\citet{Witt2009} attributes both emission and absorption in H$\alpha$ to the same structure,
which is a jet launched from the accretion disk of the companion. 
P~Cyg-like profile forms when the blue-shifted lobe
becomes projected against the face of the giant, which is impossible for the red-shifted lobe
and which is therefore always observed in emission. The jet launched
in a precessing accretion disk around the companion of
HD\,44179 may have created the Red Rectangle nebula \citep{Velazquez2011}.
This model can be adopted for BD+46$\degr$442 (Fig. \ref{fig_cart}),
with one caveat. Red Rectangle is a peculiar system,
because the optical flux is dominated by scattering as our line of
sight is in the orbital plane of the binary, allowing to see both jets
at a range of inclinations. For BD+46$\degr$442, on the contrary, the smaller IR excess indicates
that we are observing the star directly, along the line-of-sight which
is not in the orbital plane. The small line-of-sight reddening shows
that our aspect angle towards the system is not through the puffed-up
circumbinary disk.
The blue absorption as seen at conjunction is created by the
blue-shifted jet which is projected on the giant only in this specific
orbital phase. Continuum photons coming from the giant will be scattered
outside the line-of-sight by H$\alpha$ resonant scattering on high
velocity hydrogen atoms in the jet. The Doppler shift of 300 km~s$^{\mathrm{-1}}$
corresponds to the line-of-sight velocity component and needs to be
deprojected to obtain the outflow velocity of the jet. As we do not constrain the
jet opening angle, nor the angle between the jet axis of symmetry and the
orbital plane, the deprojected velocity is difficult to quantify.
The static double-peak emission still requires a different origin,
as discussed in the previous section.

The existence of jets has been strongly advocated based on other independent arguments
in e.g., a $\beta$ Lyr system \citep{Harmanec2002, Ak2007},
a proto-type of the massive Roche-overflow binaries (of which W Ser is a sub-class).
Jets have been invoked to explain bipolar planetary and proto-planetary nebulae (PPNe),
including the Red Rectangle \citep{Cohen2004,Velazquez2011}.
\citet{Sahai2002} detected a proper motion of jets in the PPNe Henize 2-90
with velocity of a few hundred km~s$^{\mathrm{-1}}$, which is consistent with the Keplerian velocity
at the surface of a low- to intermediate-mass MS star. 
Finally, jets have been also mentioned in symbiotics, in regard to
symmetric bumps at $\pm1000$ km~s$^{\mathrm{-1}}$ appearing in outbursts \citep{Skopal2009},
though they have never been invoked to explain the double-peaked profile of the quiescent state.
It is difficult to include jets in the hydrodynamic calculations due to the supersonic velocities,
or in the radiative transfer calculations. Nevertheless, the possibility of
jet contribution into H$\alpha$ and H$\beta$ formation was demonstrated
in several studies \citep{Budaj2004, Arrieta2003}.

%

\subsection{Gas stream}

Further phenomenological similarity between BD+46$\degr$442
and the mass-exchanging binaries can be seen in the non-photospheric ``shell''
components of metal lines. The narrow features, mostly in absorption, discussed in BD+46$\degr$442
(such as the \ion{Fe}{II} (42) and the near-IR \ion{Ca}{II} triplets), have been first noted in fast rotating 
hot stars \citep{Slettebak1986}.
They originate from the meta-stable levels at 2$-$4eV above the ground state, and
therefore must form in a low-density, but hot (several kK) circumstellar medium. 
In cooler stars, shell components have been mostly studied in the
resonance \ion{Ca}{II} \rm{H\&K} and \ion{Na}{I} \rm{D} lines, but these transitions are prone to interstellar
contamination.
Because of their regular behaviour with the orbital phase in BD+46$\degr$442, we again
consider them to originate in the binary flows, rather than
in some clumps in the circumbinary disk (an explanation proposed for the protoplanetary disks, see e.g. \citet{Mora2004}).
The behaviour of these features appears to be very diverse from system to system,
so it is not surprising that they were ascribed to a whole range of
accretion-related structures: disk, disk-wind collision shocks, jets, streams, etc.
\citep[e.g.][]{Andersen88, Plavec1988, Weiland1995, Miller2007, Sudar2011, Mennickent2010,
Harmanec2002, Quiroga2002}.

We found a particularly good agreement between
the behaviour of the metal absorptions in BD+46$\degr$442 with
\ion{He}{I} lines in a W Ser-type star RY Sct ($P_{\mathrm{orb}}=11^{\rm d}$, orbital separation $\sim$0.2 AU),
as described in \citet{Grundstrom2007}.
To explain these features, one needs gas projected on the giant over at least half of the period
and moving with a slower velocity, for example a circumbinary Keplerian disk.
Furthermore, this ring must have an asymmetric density distribution to explain unequal strength
of the absorption in elongations.
\citet{Grundstrom2007} proposed that this structure could in fact be
a plume on the trailing side of the giant,
an outflow from the L2 point that extends beyond the orbital plane and warps
around the system (Fig. \ref{fig_cart}). It explains why the central absorption
is clearly visible when the giant recedes from us, but is much less pronounced in the opposite elongation.
This outflowing arm is consistently reproduced in the
hydrodynamical simulations of interactive binaries \citep{Valborro2009, Mohamed2011}.
\citet{Grundstrom2007} further speculate that the outflows through the L2 and L3 points
could explain the formation of the large nebula (R$\sim$2000 AU)
surrounding RY Sct. This scenario brings an intriguing possibility
of the current formation of the circumbinary disk around BD+46$\degr$442.
Finally, in a recent study, \citet{Thomas2011} reports on the identical
behaviour of some metal lines, including \ion{Fe}{II} 4924 \AA\,, in HD 44179.
Unlike \citet{Grundstrom2007}, they interpret it in terms of the photospheric line asymmetries
due to matter outflowing from the giant near periastron and falling back near apastron.
This model, however, is questionable for BD+46$\degr$442 
due its low eccentricity. Obviously, observations of more systems with
different eccentricities and orbit orientation are needed
to test these hypotheses.

Another manifestation of the circumstellar gas around BD+46$\degr$442 are two weak emission
lines of the \ion{Fe}{I}(12) multiplet in the far red region of the optical spectrum.
These emission lines have been reported
at least in two other disk post-AGB sources, 89 Her \citep[a binary
with $P_{\mathrm{orb}}=288^{\rm d}$, $a\sin i=0.1$ AU, $i=15\degr$,  $e=$0.2,
and a resolved bipolar \rm{CO} outflow, see][]{Waters1993b,Bujarrabal2007}
and a candidate RV Tau star QY Sge \citep{Rao2002}.
In 89 Her these lines are narrow (FWHM$<$10 km~s$^{\mathrm{-1}}$) and single-peaked, while in QY Ser
and BD+46$\degr$442 they are much broader (FWHM$\sim$40 km~s$^{\mathrm{-1}}$) and
perhaps even double-peaked. Unfortunately, the latter is difficult to measure accurately for BD+46$\degr$442
due to contamination with telluric absorption.
\citet{Rao2002} proposed that this emission
originates in the bipolar wind from the system, in particular, from
the gas within the inner cavity of the circumbinary disk that re-emits starlight scattered by
the disk walls. Assuming a total stellar mass for BD+46$\degr$442 of 1.5 $M_{\odot}$ and
a circumbinary gas disk with Keplerian rotation,
the HWHM$=$20 km~s$^{\mathrm{-1}}$ indicates a distance of 3.3 AU.
Given that the inner radius of the dusty disk is at 7 AU, as deduced from the SED fitting,
this gas can indeed be located within the inner disk hole.
\citet{Rao2002} also explained broader emissions in H$\alpha$ and the \ion{Na}{I} D lines
as originating in the same wind, only further out from the disk plane where the wind has been accelerated.
This scenario would explain the lack of the RV variations in the emission lines of BD+46$\degr$442, including H$\alpha$.
The remaining differences between 89 Her and QY Sge with BD+46$\degr$442
could be due to different inclination angles.
To constrain the latter for BD+46$\degr$442, an accurate light-curve is needed.

The next obvious step to confirm the association of all these circumstellar features with binarity,
would be a systematic comparison of disk sources to the sources with spherical shells
(as inferred from the SED, e.g. \citet{VanArle2011}), as the latter presumably harbor single stars.  

\section{Summary}

We obtained 60 echelle spectra over a period of 3.5 years for BD+46$\degr$442,
a poorly-studied high-galactic giant with a dusty disk.
We derive the following photospheric parameters:
$T_{\mathrm{eff}}=6250 \pm 250\,$ K, $\log g=1.5 \pm 0.5$, and an average metallicity $\mathrm{[M/H]}=-0.7 \pm 0.2$,
without a strong depletion pattern.
The enhanced abundance of $\alpha$ elements is characteristic
of the original composition of a thick-disk low-mass star, consistent with a post-AGB
interpretation.
The observed large amplitude of the RV variations and a lack of a strong variability
in $T_{\mathrm{eff}}$  argue against pulsations.
We therefore attribute the RV variations in BD+46$\degr$442 to binarity (with a tentative detection
of a companion spectrum in the CCF). We find the following values for the orbital parameters:
an orbital period 140.8 days, an eccentricity 0.08, and separation $<$1 AU.
This adds to the several dozen post-AGB disk binaries with known orbital parameters.
The orbital period falls on the short side of the period distribution
of the other orbits.

Time-resolved spectroscopy allowed to detect gas streams in BD+46$\degr$442,
indicating that it is an interactive binary.
H$\alpha$ (and to a smaller extent higher Balmer lines and the \ion{Na}{I} D)
are found to alternate between double-peaked emission, which is
characteristic of a Keplerian disk, and a P~Cyg-like profile,
which is characteristic of an outflow. The blue absorption reaches values up to 300 km~s$^{\mathrm{-1}}$
and develops only around the giant's superior conjunction.
We suggest it is due to a jet that originates in the accretion disk around the companion.
In addition, strong metal lines display central or slightly blue-shifted absorption
inside the photospheric profiles. Similar features are also observed in some massive interactive binaries and
likely in several other post-AGB disk sources (like in the famous Red Rectangle).

The discovery of mass transfer and, possibly, accretion-driven jets in post-AGB binaries
opens a new avenue to explore some of the long-standing puzzles of these objects,
such as a variable mass-loss rate, a long life-time of the circumbinary disks,
a lack of extended dusty nebulae \citep{Siodomiak2008, Lagadec2011},
and the bipolar structure of the gaseous outflows \citep{Bujarrabal2007}.
By uncovering more systems like this in the course of our spectroscopic survey,
aided by complimentary observations,
we hope to pin-point the exact geometrical structures responsible for the various non-photospheric features.
This will enable to construct realistic hydrodynamical models,
providing a more solid basis for the theory of binary evolution.


\begin{acknowledgements}

We would like to thank Dr. V. Kovtyukh for sharing his atomic data,
Dr. P. Degroote for his SED tool, Dr. L. Winter for using her de-reddening script,
the numerous observers who obtained the spectra, and the anonymous referee for the useful suggestions.
The HERMES project and team acknowledge support from the Fund for Scientific
Research of Flanders, Belgium (FWO), support from the Research Council of
K.U.Leuven (Belgium), support from the Fonds de la Recherche Scientifique,
Belgium (FNRS), from the Royal Observatory of Belgium and from the
Landessternwarte Tautenburg (Germany).
This publication makes use of data products from AKARI, a JAXA project with the participation of ESA,
and from the Wide-field Infrared Survey Explorer,
which is a joint project of the University of California, Los Angeles,
and the Jet Propulsion Laboratory/California Institute of Technology,
funded by the National Aeronautics and Space Administration.

\end{acknowledgements}

\bibliography{gorlova_bdpl46_3}

\appendix
\section{Representative line profiles}

\begin{figure*}
   \centering
   \includegraphics[width=\textwidth]{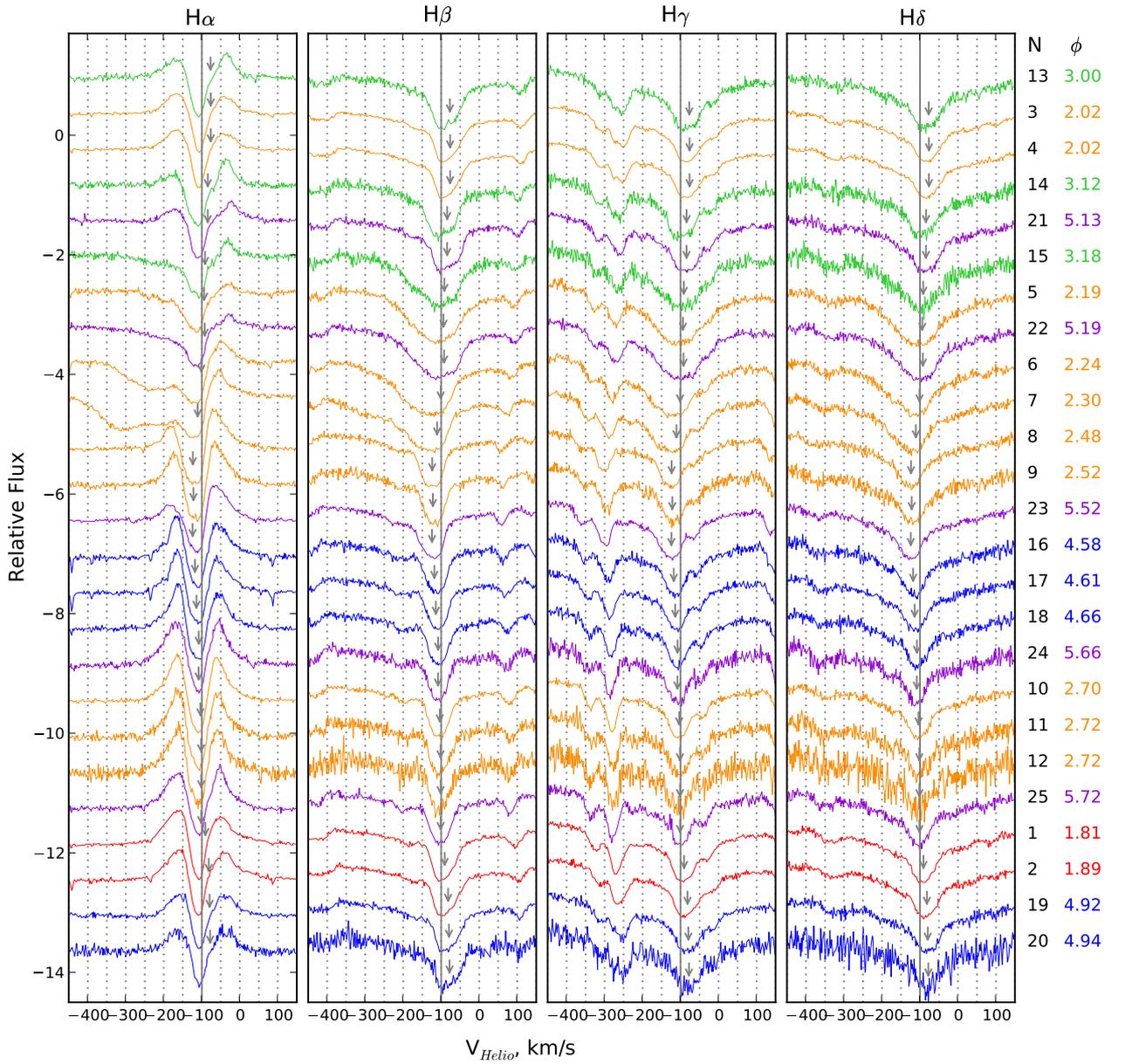}
      \caption{
Balmer lines as a function of the RV phase ($\phi=0$ corresponds to
the maximum redshift).
Different colours denote different orbital cycles.
A solid vertical line marks our systemic velocity of -98.9 km~s$^{\mathrm{-1}}$,
while dotted lines mark 50 km~s$^{\mathrm{-1}}$ intervals from it.
The arrows mark the photospheric velocity according to the CCF.
}
         \label{fig_Balmer}
   \end{figure*}

Fig. \ref{fig_Balmer} shows H$\alpha$ and three other Balmer lines in BD+46$\degr$442
ordered according to the orbital phase, where each of the five orbital cycles (covering the
first 1.5 years of observations) is designated with a different colour.
A corelation with the orbital phase is obvious.
The central absorption in H$\alpha$ is not always aligned with the
photospheric velocity, which means that the profile can not be represented
by a simple superposition of a broad emission with a photospheric absorption.
Near giant's inferior conjunction ($\phi = 0.75$) H$\alpha$ exhibits
a double-peak emission, while near
superior conjunction ($\phi = 0.25$) the blue peak is replaced with
an extended blue absorption, as can also bee seen in the higher Balmer lines.

\begin{figure*}
   \centering
   \includegraphics[width=\textwidth]{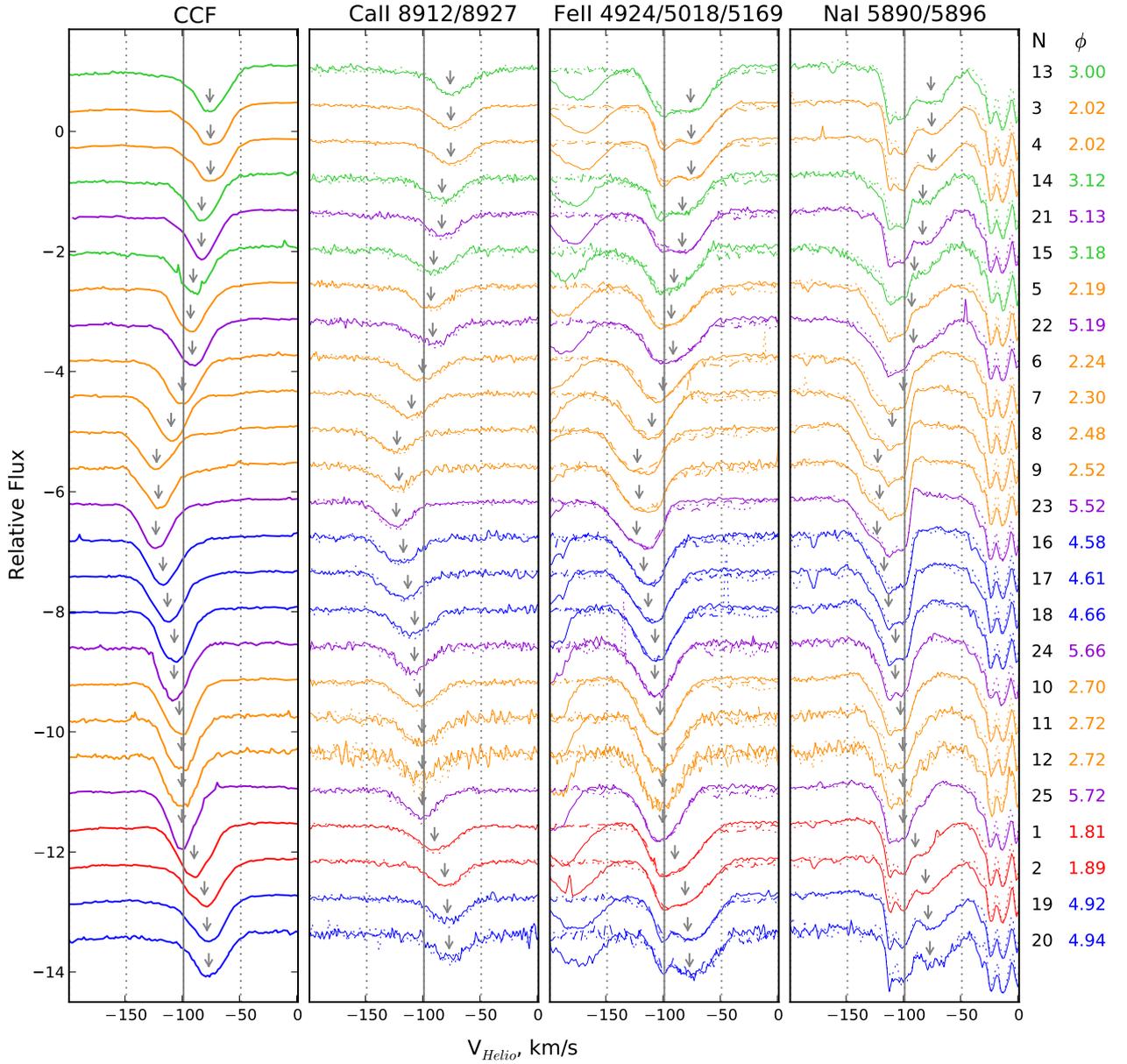}
   \caption{
Same as in Fig.\ref{fig_Balmer}, only for the CCF and the representative metal lines of different strengths:
\ion{Ca}{II} 4d-4f doublet, \ion{Fe}{II} (42) triplet, and the \ion{Na}{I} D doublet.
In each panel different line styles (solid, dashed, dotted) designate
different members of the same multiplet.
}
         \label{fig_multiplets}
   \end{figure*}

Fig. \ref{fig_multiplets} shows the behaviour of some metal lines and the CCF.
The CCF represents the behaviour of the majority of the photospheric lines, that
are symmetric and move with the orbital velocity.  Low-excitation ($\chi_{\mathrm{low}}=$0--3 eV),
strong (EW $>$300 m\AA) lines, on the other hand,
show additional circumstellar and interstellar (in \ion{Na}{I} D) components, particularly between $\phi=0.9-1.1$.

%

\end{document}